\documentclass[a4paper,11pt]{article}
\usepackage{pos}
\usepackage{wrapfig}

\title{The Terzina payload on board the NUSES space mission}

\author*[a]{Teresa Montaruli for the NUSES Collaboration}

\affiliation[a]{D\'epartement de Physique Nucléaire et Corpusculaire, Facult\'e de Science, Universit\'e de Gen\'eve, 24 Quai Ernest-Ansermet, 1205, Switzerland}

\emailAdd{teresa.montaruli@unige.ch}

\abstract{The Terzina payload, onboard the NUSES space mission is being built in collaboration with TAS-I by GSSI, INFN and the University of Geneva. It is a Cherenkov Schmidt-Cassegrain  compact telescope with an effective focal length of 925~mm and a camera focal assembly composed of 640 pixels (16 vertically and 40 horizontally) organized in $8\times 8$ tiles produced by FBK with sensitive area of $2.73 \times 2.64$~mm$^2$. We will illustrate the performance for the signal of cosmic rays beyond a threshold of a few hundred of PeV. Understanding the operation of SiPMs in space with almost direct exposure to solar and trapped protons and electrons, defining the mitigation strategy for the increase of DCR of exposed silicon, the characterisation of the luminous backgrounds, the data acquisition strategy for this payload, the maximization of the effective exposure to the atmospheric showers induced by the signal of neutrinos and cosmic rays are the challenges. At this conference, we describe how they are being addressed.}

\ConferenceLogo{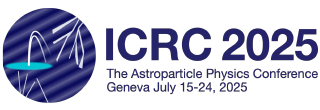}

\FullConference{39th International Cosmic Ray Conference (ICRC2025)\\
 15–24 July 2025\\
Geneva, Switzerland\\}

\begin{document}
\maketitle

\section{Terzina onboard NUSES satellite}

The detections of cosmic neutrinos by IceCube \cite{IceCube:2018cha} and gravitational waves by LIGO and Virgo \cite{TheLIGOScientific:2017qsa} have revolutionized the field of multi-messenger astrophysics. This is the domain of collaborative observations of many observatories and experiments in all bands of the electromagnetic spectrum with these elusive and newly discovered messengers. These bring information from regions where photons are absorbed and on new cosmic sources.
To achieve cosmic neutrino detection, initial efforts were conducted in underground environments, such as mountain and mines. The successful detection required a cubic-kilometer neutrino telescope at more than 2 km depths in the Antarctic ice \cite{IceCube_diffuse} and at more than 3 km in the sea \cite{KM3NeT_event}.

As fluxes from Nature follow naturally decreasing power law spectra, the scarcity of source luminosity at higher and higher energies, where the cosmic signal dominates the atmospheric backgrounds of muons and neutrinos, pushes cosmic ray and neutrino searches towards new cost-effective technologies, like radio detection or space-based and high-altitude balloons. These profit from large exposures of the atmosphere which is used as a large radiator for Cherenkov or for fluorescence emission from nitrogen atoms de-excitation.
From these platforms the visible target mass of the Earth for neutrino interactions is suitable for fluxes that in a cubic-kilometer detector like IceCube produce $\sim 10$~events per year beyond 50~TeV. Also ultra-high energy cosmic rays (UHECRs) can be measured in regions between about 1-500 PeV from balloons and beyond 50~PeV from space for field of views of similar dimension than Terzina.
This region is quite critical, and still poorly explored, for the study of the spectrum and composition of cosmic rays to understand the transition from the highest energy tail of the galactic flux and the onset of the extragalactic one.

Several projects are aiming to detect for the first time the fluorescence and Cherenkov light produced by cosmic particles in the atmosphere from space. POEMMA is a project with a fleet of two 4-ton satellites on a 525~km orbit with $28.5^\circ$ inclination. The satellites can slew at a distance of $\sim 300$~km to cover a surface on Earth of about 500~km$^2$ for UHECR detection using fluorescence beyond about 20 EeV or at 30 km distance to search for Cherenkov emission from the limb from the showers induced by earth-skimming $\nu_\tau$'s and $\nu_\mu$'s beyond 20~PeV from a $\sim2000$~km$^2$ area.
A lower cost alternative is a fleet of middle-size satellites, such as the proposed CHANT \cite{CHANT}. Other projects aim at the detection of induced showers from high atmosphere, as the Super Pressure Balloon 2 (EUSO-SPB2) \cite{SPB2}, which unfortunately operated for a few hours, and the future Super Pressure Balloon with Radio (PBR) \cite{PBR}, which will circle over the Southern Ocean for 50 days. 
PBR \cite{PBR} is a 1.1~m aperture Schmidt telescope similar to the POEMMA design with 2 a fluorescence and Cherenkov cameras in its focal surface. This last has 2048 $3\times 3$~mm$^2$ silicon photomultipliers (SiPMs) providing a FoV of $12 \times 6$ deg$^2$, covering a spectral range of 320-900~nm. The advantage offered by high-altitude inside the atmosphere is the possibility to carry larger payloads and to be closer ($\sim10-100$~m) to the showers, thus receiving a larger amount of signal photons than from space, but a long duration of missions is a challenge. 

NUSES is a space-based mission financed by the Italian Ministry and conducted by the GSSI, INFN institutions and the UniGe. It will clarify the concept of a larger mission.
NUSES is a satellite with 2 innovative payloads dedicated to the study of cosmic radiation and the Sun-Earth environment, named Terzina and Ziré \cite{DeMitri:2023nml}.
Here we focus on Terzina, a technological pathfinder for UHECR studies and earth-skimming astrophysical neutrinos beyond 50~PeV for future larger missions, exampled of which are POEMMA \cite{POEMMA} or \cite{CHANT}, while Ziré is presented in \cite{zire}.  

\begin{figure}[t!]
    \centering
    \includegraphics[height=5cm]{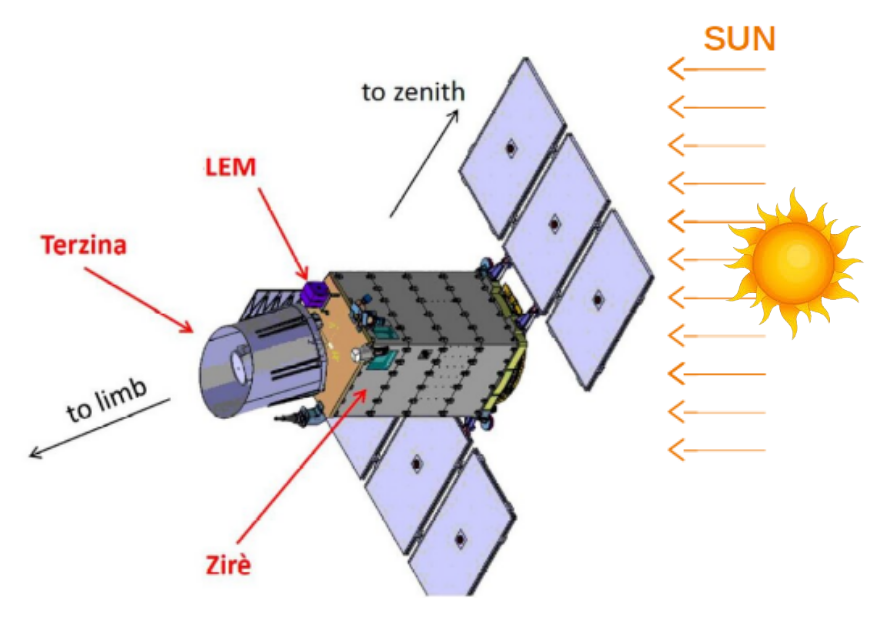}
    \includegraphics[height=5cm]{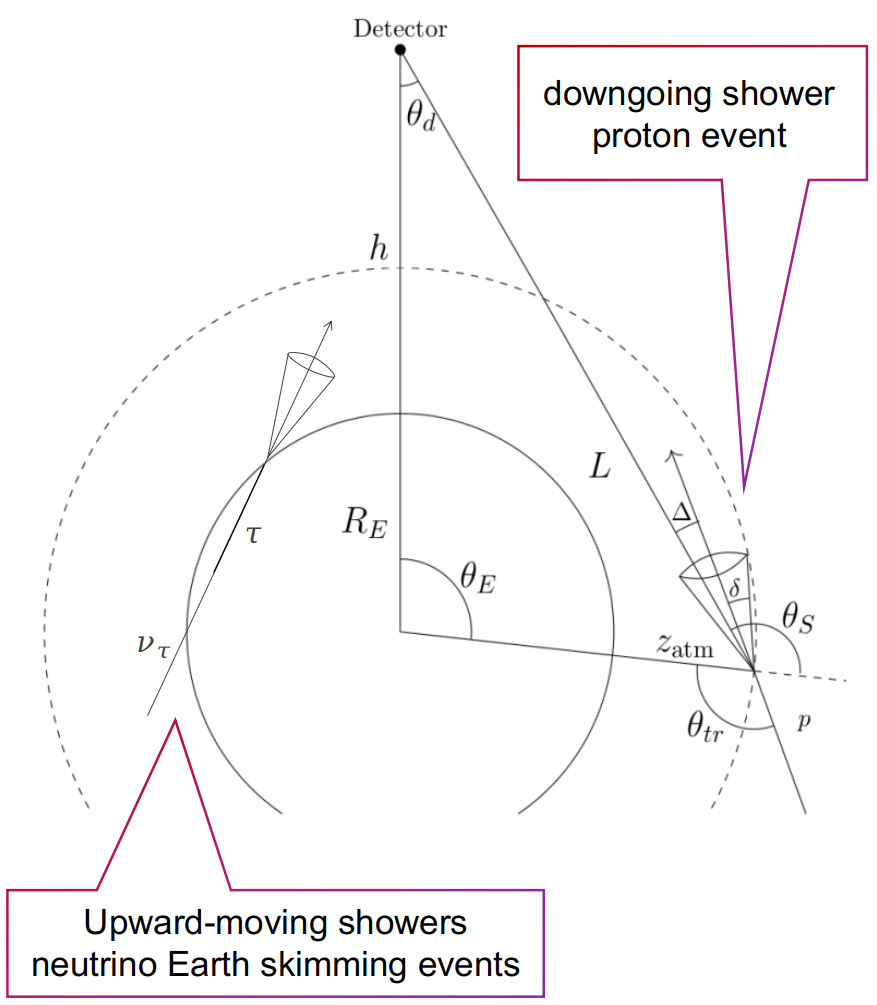}
    \caption{[Left]: the general scheme of the NUSES satellite design. [Right]: Target events of the Terzina Cherenkov telescope, which points at the limb: on the right a proton-induced shower from the atmosphere above the limb and on the left a tau neutrino-induced shower emerging from the Earth crust. The scheme is adapted from~\cite{PhysRevD.104.063029}. Both showers are directed towards the satellite and are more than 2000~km away from it.}
    \label{fig1:TerzinaTarget}
\end{figure}

The NUSES mission will last about 3 years after the launch, currently foreseen at in 2027. The satellite will be in a LEO orbit at 550~km of altitude at beginning of life (BoL) and about 525 km at end of life (EoL), inclination $97.6^\circ$ and LTAN 18:00:00, in a sun-synchronous orbit as it requires a dark atmosphere for the flebile flashes of blue light detection lasting typically $\sim10$~ns (see Fig.~\ref{fig1:TerzinaTarget}). 
The pointing accuracy ensured by the platform is $\sim 0.1^\circ$.

Terzina will detect for the first time the Cherenkov light produced by cosmic particles in the atmosphere from space, to test the performance of a SiPM-based camera, and mitigation strategies to radiation damage, and to measure the luminous pollution to the Cherenkov signal.
Differently to the foreseen POLAR-2 mission \cite{2023ExA....55..343M}, in the case of Terzina the photon signal in the SiPMs is not amplified by a protective layer of scintillators.

In this proceeding, the description of the Terzina geometry, its optical design and camera, some results of the full Monte Carlo (MC) concerning UHECRs are illustrated. Other contributions at this conference are in the posters in Ref.~\cite{Trimarelli} and Ref.~\cite{Shideh}. The results for earth-skimming neutrinos are preliminary. We expect a rate of neutrinos about $10^{-4}$ smaller than UHECRs.

\section{The Terzina components}

\subsection{The Optical Telescope Assembly (OTA)}

\begin{figure}[t!]
    \centering
    \includegraphics[height=5.5cm]{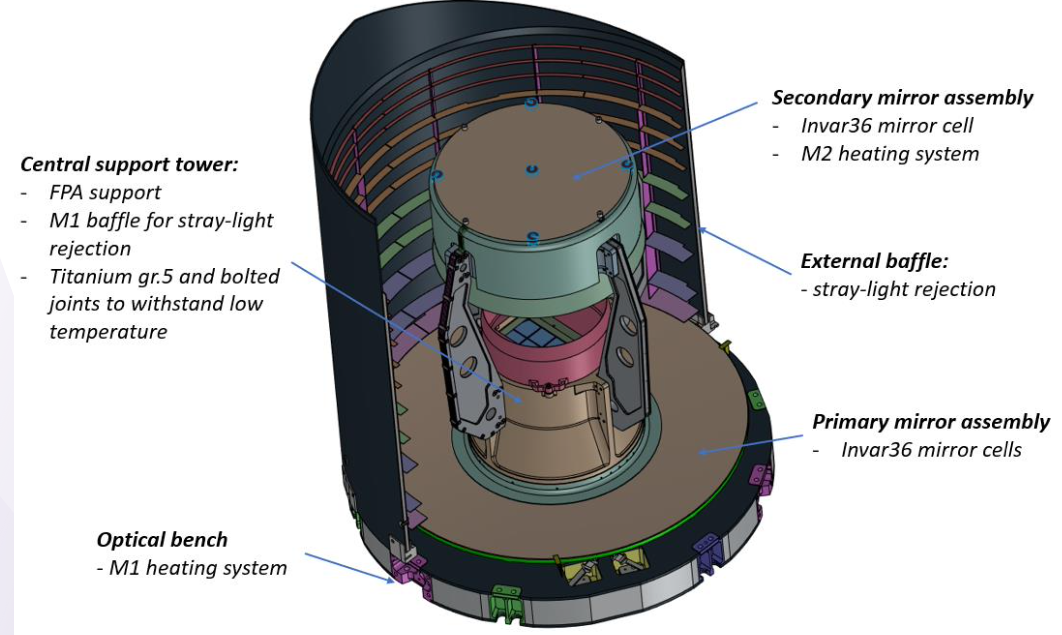}
 \includegraphics[width=0.35\textwidth]{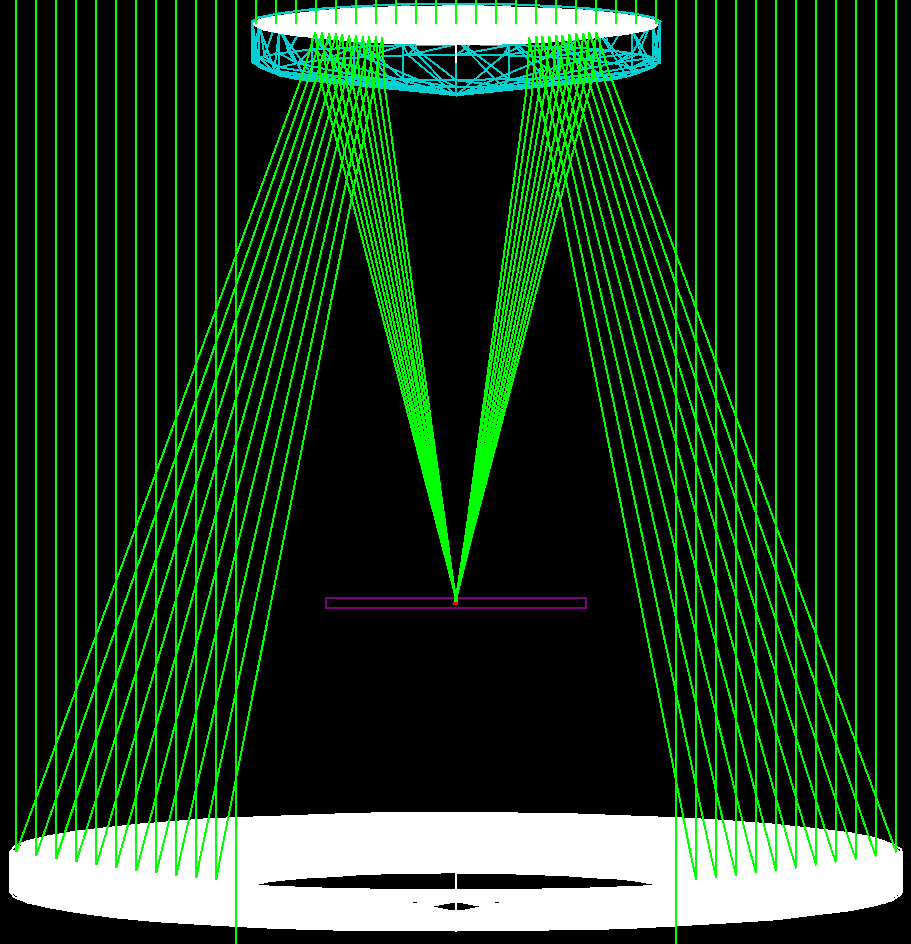}
    \caption{[Left]: Image from the CAD of Terzina also used by the Geant4 simulation, with components indicated. [Right]: The ray-tracing simulation of Geant4 showing the optical path of the light. 
}
    \label{fig:detector}
\end{figure}

The Terzina detector (see Fig.~\ref{fig:detector}-left) consists of a near-UV optical telescope with Schmidt-Cassegrain (in the Ritchev-Chrétien configuration) dual mirror system
and equivalent focal length of 925~mm, named OTA. It is being realised by the company Officina Stellare (OS), after the specifications provided by the UniGe and GSSI. The Focal Plane Assembly (FPA) is the focal plane in the round hole of the primary mirror.
The total collection area of the primary and secondary mirrors is 0.11 m$^2$, including the shadowing of the baffle and vanes holding the secondary of $\sim 8\%$, an average reflectivity of mirrors of 92\% between $300-650$~nm, a superior cut due to a filter on the secondary mirror to reduce the night sky glow (NSG). 
In this mission, the primary goal is the first detection of UHECRs while it is improbable that Terzina will observe earth-skimming neutrinos, which are a factor of $\sim 10^{-4}$ rarer than UHECRs.

\begin{figure}[t!]
    \centering
        \includegraphics[width=0.5\textwidth]{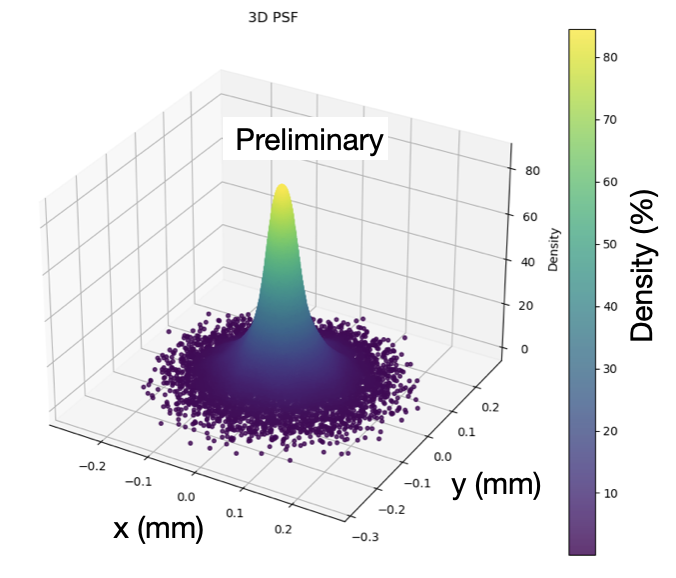}
           \includegraphics[width=0.3\textheight]{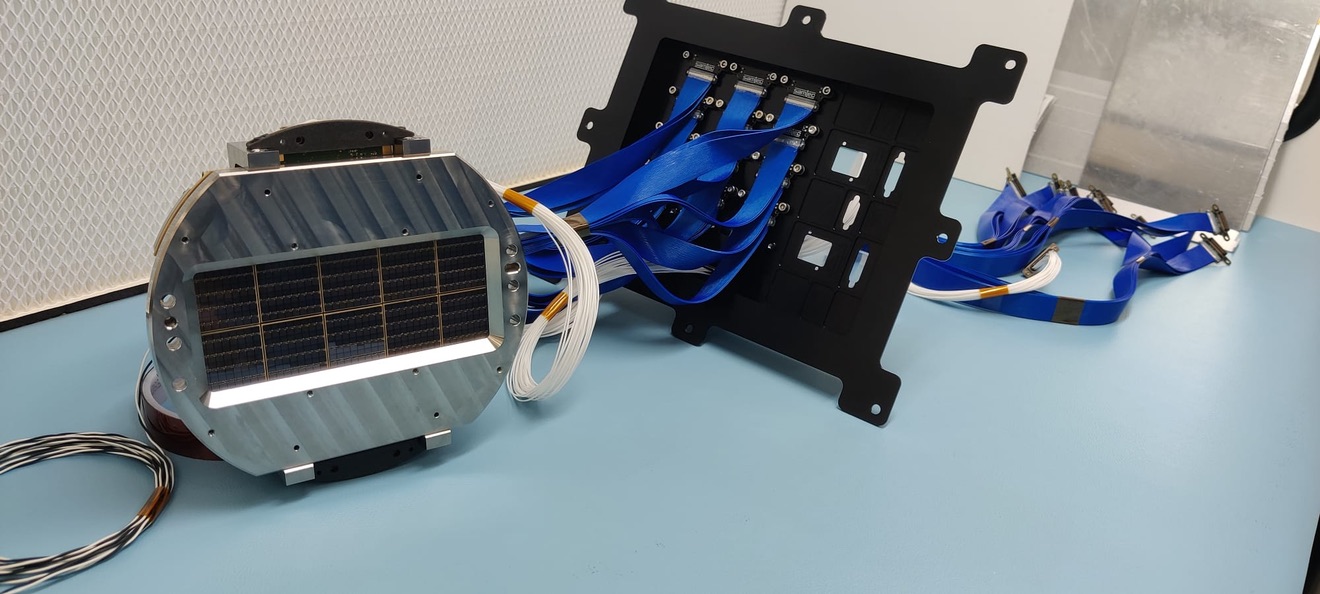}
    \caption{[Left]: The PSF of the optical system on axis. On the z axis is the percentage of simulated photons on an area of the dimensions of a pixel on the xy plane. [Right]: the FPA being cabled in the UniGe laboratory where the CITIROC electronics was  successfully tested.}
    \label{fig:raytracing}
\end{figure}

The primary hyperbolic concave mirror has a diameter of 444~mm and radius of curvature (RoC) of 1207~mm, fitting a flat FPA in a circular hole of 230~mm diameter. The secondary aspherical (convex hyperboloid) mirror has a diameter of 201.6~mm and RoC\,= -626~mm and it is located at a distance of 428~mm from the primary mirror. The secondary mirror is covered by a correcting lens of thickness 2~mm and RoC\,=\,320~mm to compensate for aberrations also due to the flatness of the FPA.
A view of the light path between the 2 mirrors to the FPA is in Fig.~\ref{fig:detector}-right.
The point spread function (PSF) of the optical system (Fig.~\ref{fig:raytracing}-left), is fully inside a pixel sensitive area of
$2.73 \times 2.64$~mm$^2$ 
and in perfect agreement between Zemax by OS and the Geant4 simulation.

\subsection{The FPA}

\begin{wrapfigure}{l}{0.6\textwidth}
    \vspace*{-11pt}
    \centering
   \includegraphics[width=0.6\textwidth]{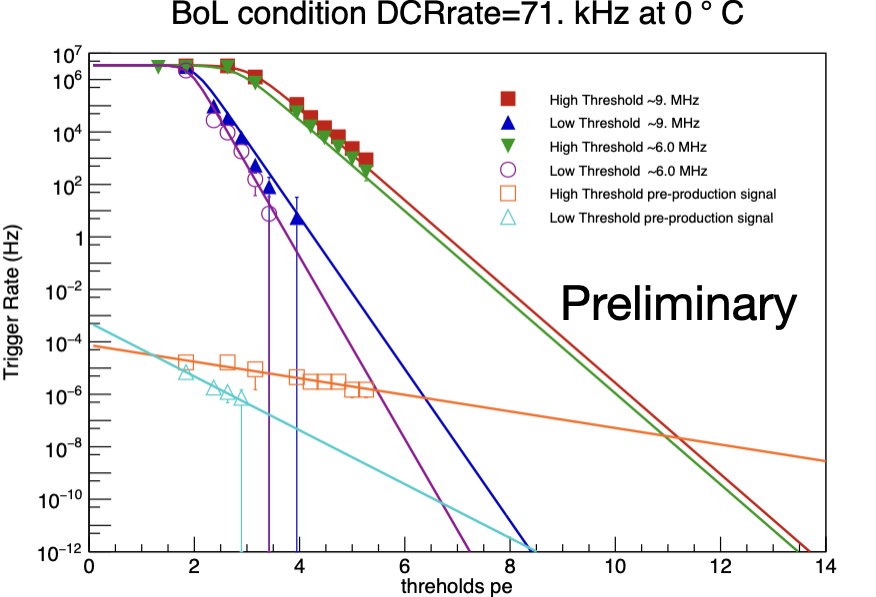}
    \caption{The trigger scan vs the threshold in p.e. Upper curves are for a large noise of 7.6 MHz (when Terzina points towards middle-size cities) for LT and HT. The lower curves are for the UHECR signal with the cosmic ray flux in \cite{PAO} for HT and LT. }
        \label{fig:trigger}
\end{wrapfigure}

The camera plane is located at 145~mm from the primary mirror. It includes 10 tiles of SiPM arrays produced for the mission by FBK of with $8\times 8$ pixels of $3\times3$~mm$^2$ (640 pixels in total) on 2 rows of 5 tiles. The total area of the tiles with their packaging within the metal frame seen in Fig.~\ref{fig:trigger}-left is $131.3 \times 58.2$~mm$^2$ and the sensitive area of a single channel is $2.63 \times 2.74$~mm$^2$, so due to the packaging the sensitive area is $\sim 60\%$. The field of view (FoV) seen by each pixel is $0.18 \times 0.16$~deg$^2$ so that the total FoV is $7.2 \times 2.56$ deg$^2$.
As the optical system inverts images, the upper part of the camera will measure the light background from the Earth and possibly rare neutrinos, and the lower part the UHECRs. The FPA under tests is shown in Fig.~\ref{fig:raytracing}-right.

The electronic boards have been developed by Nuclear Instruments and are based on the \href{https://www.caen.it/products/citiroc-1a/}{CITIROC chip}. 
Each readout chip serves 32 channels, so two are needed per tile. CITIROC has low and high gain branches with 1:10 ratio for a wide range amplification. It will be used for energy measurements with a peak detector, with an excellent linearity of about 1\% in the dynamic range from 1 to about a $\sim 100$ p.e.. The time resolution is better than 1~ns. The analogue pulse in input to CITIROC from each sensor is $\rm FWHM \sim40$~ns and the gain factor (amplitude of the signal in mV/number of photoectrons - p.e.) is 3.8. CITIROC measures only the amplitude of the signal and not the full waveform, from which the visible energy can be derived with an appropriate MC study.

There will be 2 triggers: one with a low threshold (LT) that will require the coincidence inside a programmable window (e.g. 20~ns as used in the current simulation) of 2 (and in addition 3 channels will be implemented too); and one with high threshold (HT) with 1 channels beyond a higher threshold. The trigger scan obtained varying the number of photoelectron threshold is shown for the two thresholds in Fig.~\ref{fig:trigger}, where we show the signal and a 7.5~MHz noise. 
The noise comes mainly from the DCR and NSG, as detailed below.

\subsection{The tile's characterization and their radiation hardness}

\begin{figure}[t!]
    \centering
    \includegraphics[width=1.\textwidth]{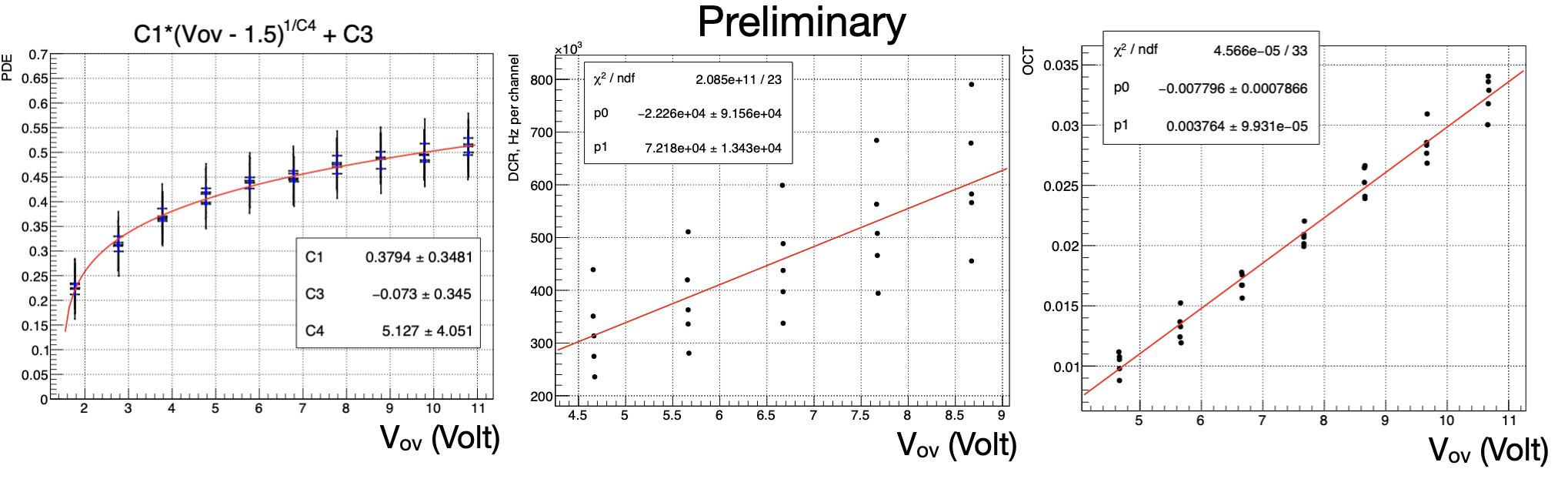}
     \caption{[Left]: The PDE; [Middle] DCR [Right]: OCT vs over-voltage of a few channels of a tile.}
   \label{fig:characterisation}
\end{figure}

The tiles were built with the constraints to have a micro-cell of 30 $\mu$m to compromise between the need of a SiPM analogue signal of slightly less than 40 ns and a photodetection efficiency (PDE)$>50\%$ at 450~nm. The dark count rate (DCR) should be less than 50 kHz/mm$^2$ and the optical cross talk (OCT) of a few percent at the desired operation over-voltage of 8-10~V. 
Once produced, the tiles were characterize to verify compliance to specifications. In Fig.~\ref{fig:characterisation} we show: the PDE, the DCR and OCT vs the over-voltage for a few channels on tested preliminarily produced tile. 

In \cite{radiation,Shideh}, we determined that without annealing, the DCR will increase up to 50 MHz, while 40\% of the SiPM response can be recovered performing annealing at temperatures of $50^\circ$C and an annealing cycle of 84 hrs.

\subsection{The signal and the night sky glow}

\begin{figure}[hbt]
    \centering
    \includegraphics[width=0.32\textwidth]{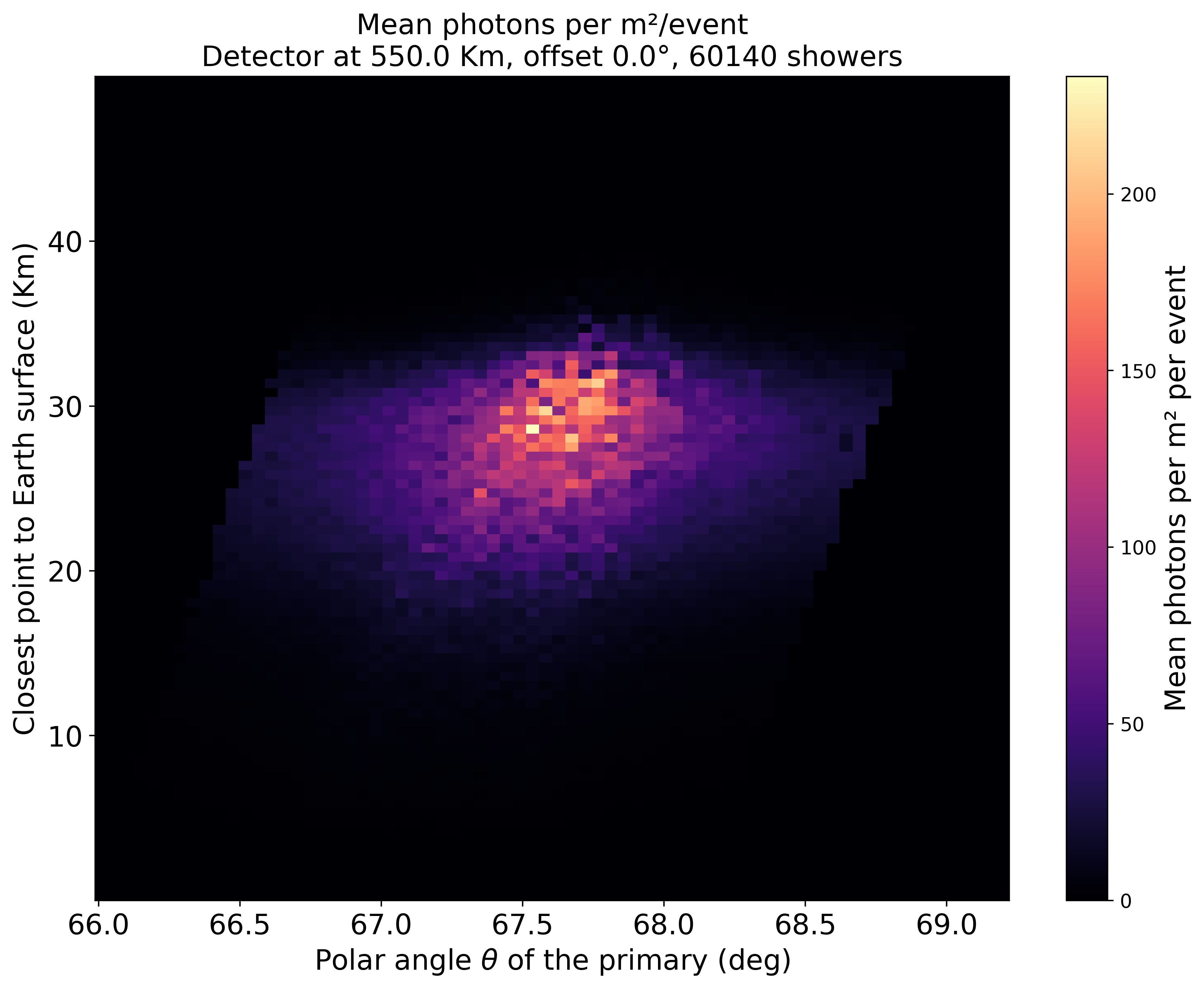}
    \includegraphics[width=0.3\textwidth]{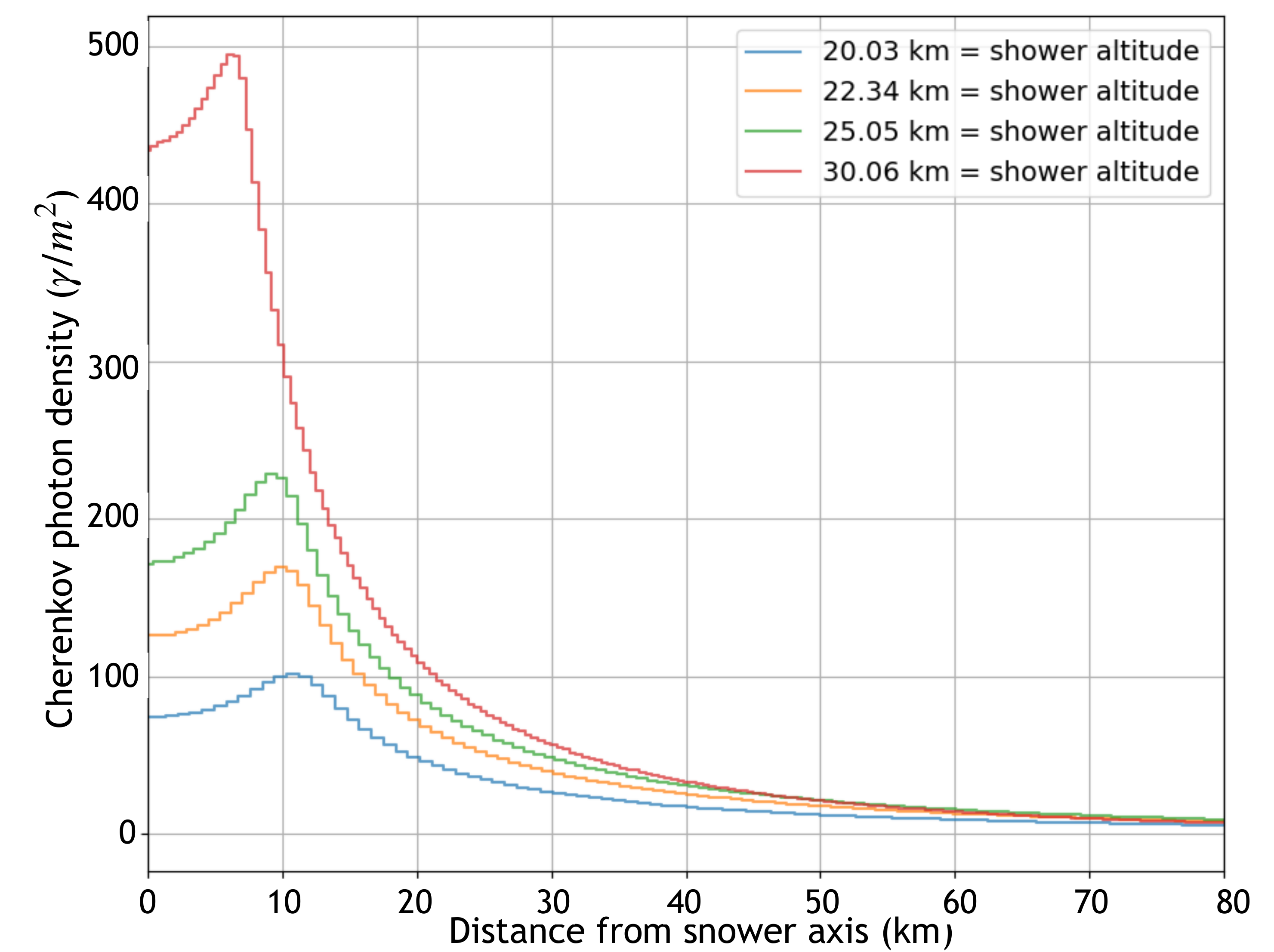}
     \includegraphics[width=0.35\textwidth]{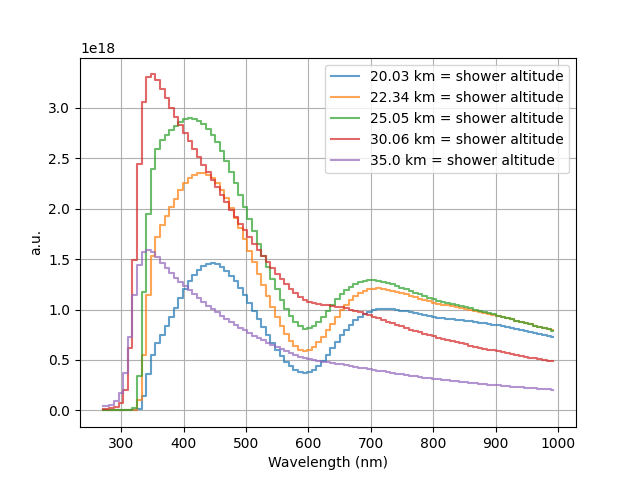}   
     \caption{[Left]: average photon number per m$^2$ for 60'000 UHECR proton-induced showers following the cosmic ray spectrum \cite{PAO} seen by Terzina at 550 km. The atmospheric layer above the limb from which photons come is $\sim 1^\circ$. [Middle]: The Cherenkov cone photon density from the axis of the telescope of the visible showers of 100 PeV produced at different altitudes. [Right:] The Cherenkov spectrum for these showers for different altitudes. The dip in the spectra is due to the ozone layer.}
   \label{fig:showers_Terzina}
\end{figure}

Terzina will see the bulk of the UHECR showers at a distance of 3500-3800~km and most of the Cherenkov photons will be seen between an altitude in the atmosphere of 25-30~km (see Fig.~\ref{fig:showers_Terzina}-left). The Cherenkov cone will have $\sim10$~km in aperture radius at that altitude and distance from the detector (Fig.~\ref{fig:showers_Terzina}-middle), so only a small fraction of the cone will be captured. To firmly reconstruct the UHECR energy and composition 3 equal satellites can determine the cone aperture.

\begin{figure}[t!]
    \centering
    \includegraphics[width=0.52\textwidth]{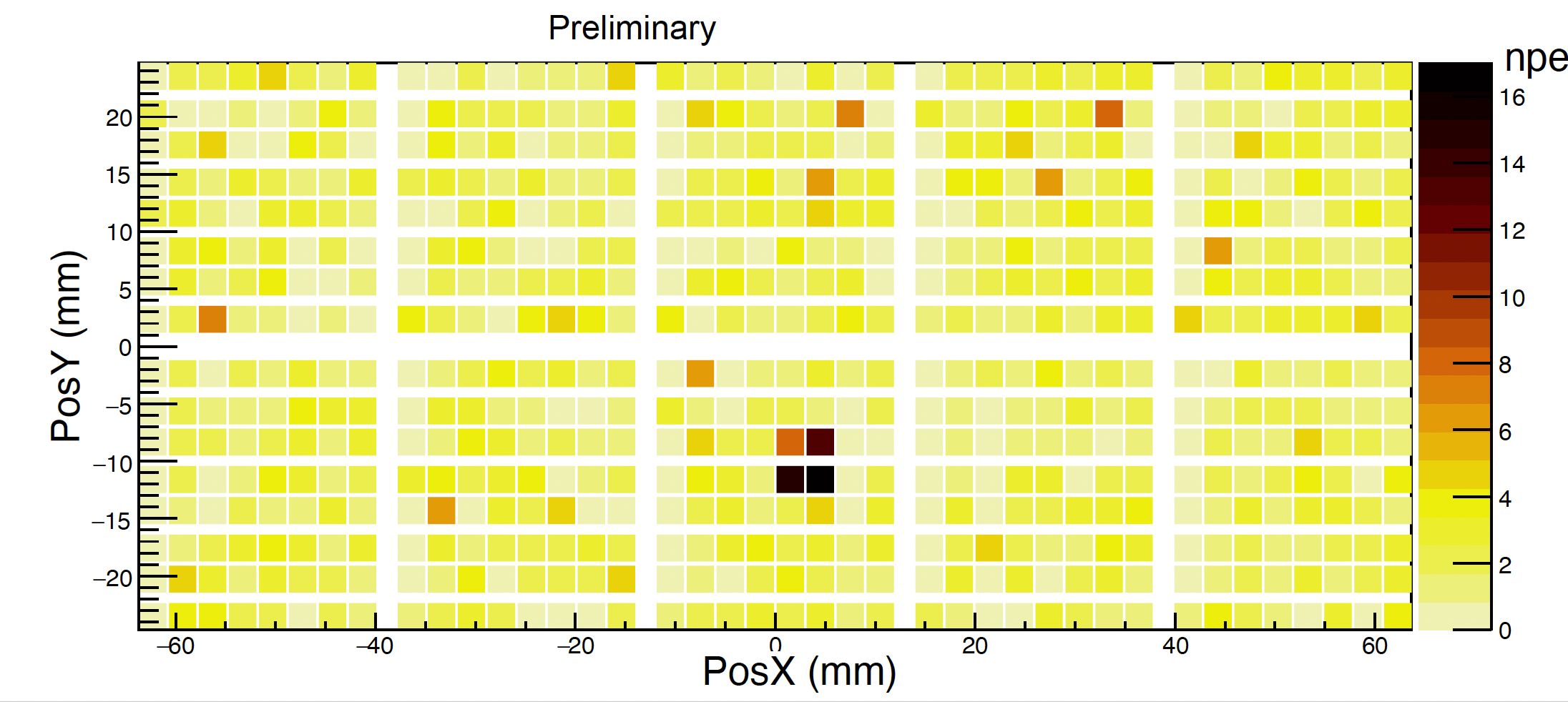}
    \includegraphics[width=0.3\textwidth]{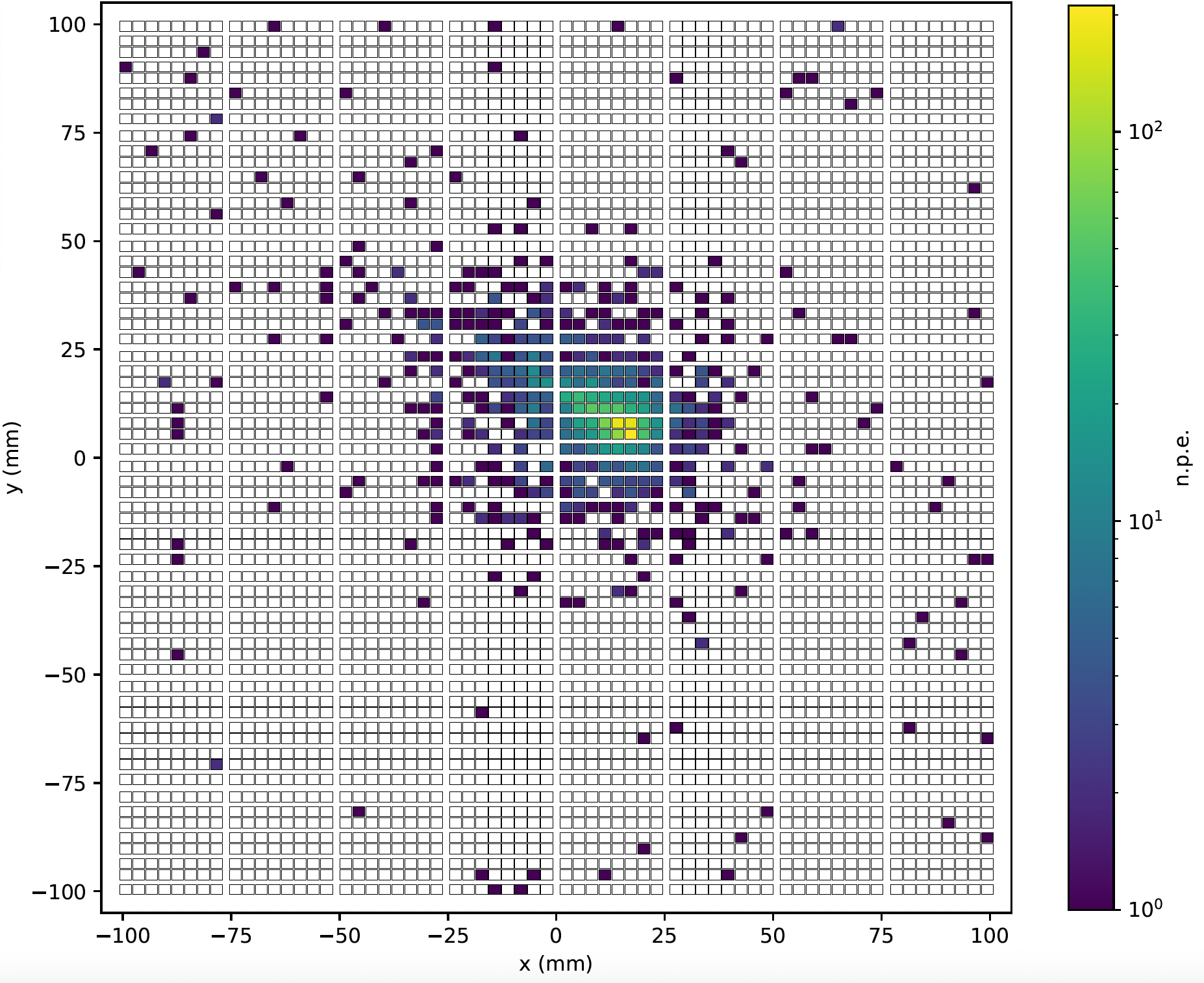}
    \caption{[Left]: a 300 PeV simulated event after LT trigger in Terzina (pixels with highest p.e. in black) (from a portion of the Cherenkov cone). [Right]: A 100 PeV simulated event for a Terzina-like detector at 30 km altitude.}
    \label{fig:events}
\end{figure}

Operating a Cherenkov detector at high altitude in the atmosphere, has the advantage to detect more photons due to proximity to showers leading to a lower energy range than from space ($\sim1$~PeV-500 ~PeV). A space-based mission with a few satellites ensures longer lifetime and can detect UHECRs up to the  ankle and GZK cutoff and reconstruct the full Cherenkov cone for composition measurement, being more cost-effective than large satellites. 
The developed simulation in the frame of NUSES is capable of handling both configurations.
Two events of are shown in Fig.~\ref{fig:events} for  Terzina-in-orbit and a similar telescope to Terzina ($\sim 1.6$ larger) at 30~km in the atmosphere.  

\begin{figure}[t!]
        \includegraphics[width=0.5\textwidth]{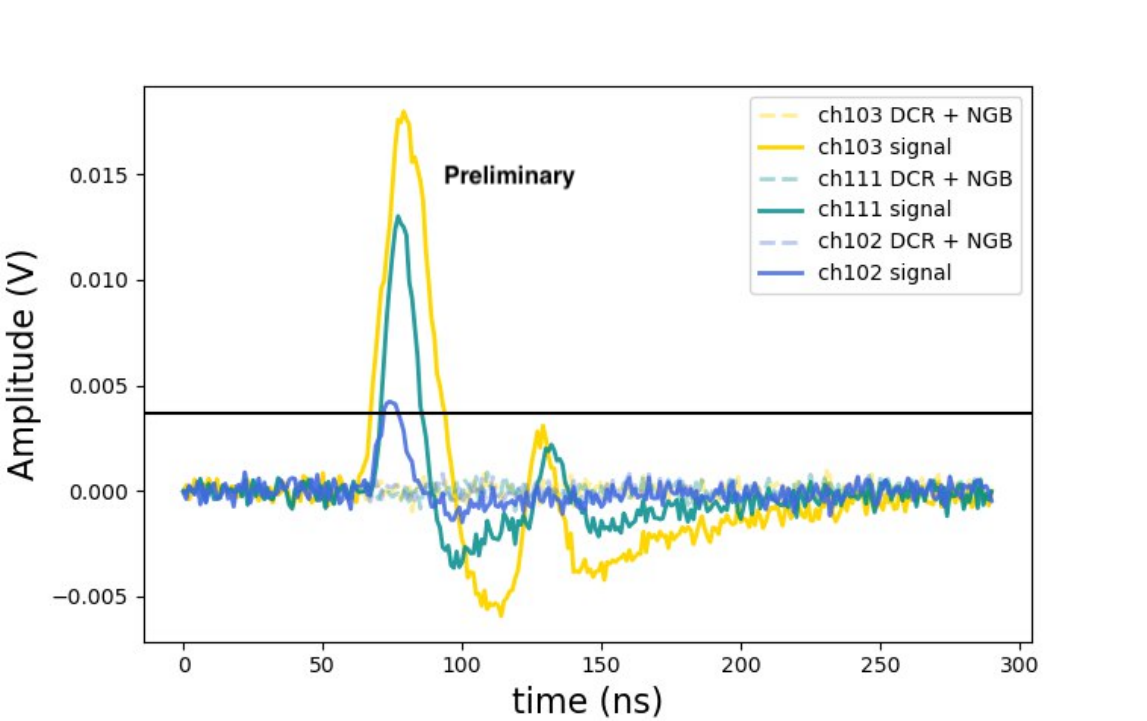} 
    \includegraphics[width=0.43\textwidth]{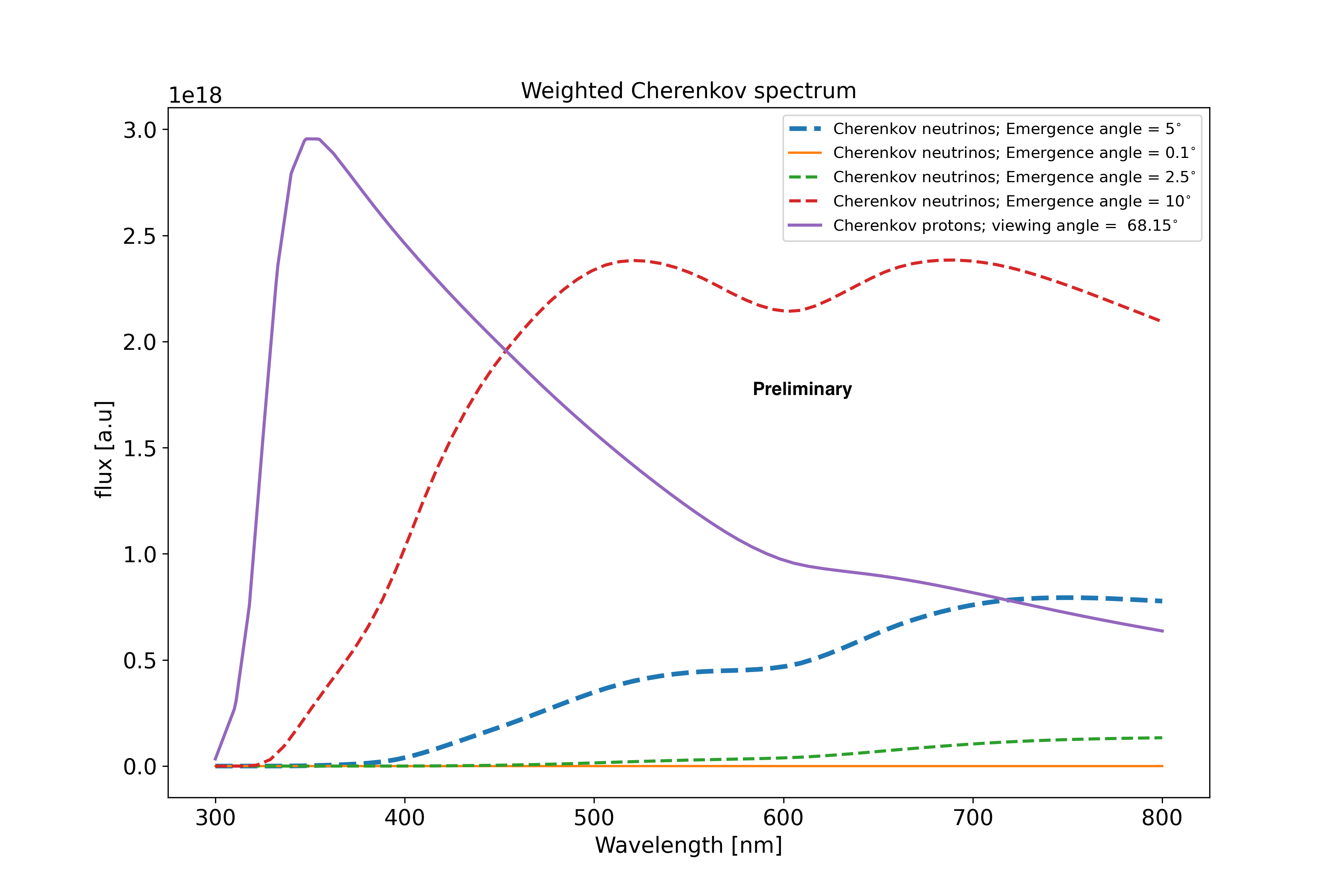} 
    \caption{[Left]: waveform noise (NSG+DCR) and signal. [Right]: wavelength spectrum of $\nu$-induced showers in arbitrary units for some emergence angles from ground and so different distances from Terzina.}
    \label{fig:signal}
\end{figure}

The waveforms of the signal of a triggered event with the LT scheme superimposed on a 10 MHz background is shown in Fig.~\ref{fig:signal}-left. CITIROC does not reconstruct the waveform (see Fig~\ref{fig:events}-right), but only measures the amplitude. For a future development, a low-power consuming ASIC will be used (see e.g. the work of the Torino-INFN group \cite{INFNTo}). 

The NSG has been calculated more precisely as it comes principally from luminous pollution from the Earth and the Moon. When the Moon will be in the FoV of Terzina (less than 10 min every year on the orbit), measurements will not be possible as rates will be too high. The ideal condition will be on the oceans with NSG rates of $<10$~kHz and on the continent in the presence of small cities $<300$~kHz. These estimates were based on the \href{https://web.archive.org/web/20051214020538/http://dmsp.ngdc.noaa.gov/html/sensors/doc_ols.html}{DMSP satellite data} as they would be seen by Terzina in its orbit.

We expect to observe 88 UHECR events/yr for a 200 kHz background (in the presence of small cities). For a firm neutrino detection a fleet or larger mission will be needed, but Terzina's upper limits will be unprecedented.

\paragraph{Acknowledgements}
NUSES is funded by the Italian Government (CIPE n. 20/2019), by the Italian Ministry of Economic Development (MISE reg. CC n. 769/2020), 
by the Italian Space Agency (CDA ASI n. 15/2022),  by the European Union NextGenerationEU under the MUR National Innovation Ecosystem  
grant ECS00000041 - VITALITY - CUP D13C21000430001, and in collaboration with the Ministry of University and  Research, MUR, under contract n. 2024-5-E.0 - CUP n. I53D24000060005.

\begin{center}
{
\large 
\bf
The NUSES Collaboration 
}
\vspace{5ex}

M.~Abdullahi$^{a,b}$, R.~Aloisio$^{a,b}$, F.~Arneodo$^{c,d}$, S.~Ashurov$^{a,b}$, U.~Atalay$^{a,b}$, F.~C.~T.~Barbato$^{a,b}$, R.~Battiston$^{e,f}$, M.~Bertaina$^{g,h}$, E.~Bissaldi$^{i,j}$, D.~Boncioli$^{k,b}$, L.~Burmistrov$^{l}$, F.~Cadoux$^{l}$, I.~Cagnoli$^{a,b}$, E.~Casilli$^{a,b}$, D.~Cortis$^{b}$, A.~Cummings$^{m}$, M.~D'Arco$^{l}$, S.~Davarpanah$^{l,z}$, I.~De~Mitri$^{a,b}$, G.~De~Robertis$^{i}$, A.~Di~Giovanni$^{a,b}$, A.~Di~Salvo$^{h}$, L.~Di~Venere$^{i}$, J.~Eser$^{n}$, Y.~Favre$^{l}$, S.~Fogliacco$^{a,b}$, G.~Fontanella$^{a,b}$, P.~Fusco$^{i,j}$, S.~Garbolino$^{h}$, F.~Gargano$^{i}$, M.~Giliberti$^{i,j}$, F.~Guarino$^{o,p}$, M.~Heller$^{l}$, T.~Ibrayev$^{c,d,q}$, R.~Iuppa$^{e,f}$, A.~Knyazev$^{c,d}$, J.~F.~Krizmanic$^{r}$, D.~Kyratzis$^{a,b}$, F.~Licciulli$^{i}$, A.~Liguori$^{i,j}$, F.~Loparco$^{i,j}$, L.~Lorusso$^{i,j}$, M.~Mariotti$^{s,t}$, M.~N.~Mazziotta$^{i}$, M.~Mese$^{o,p}$, M.~Mignone$^{g,h}$, T.~Montaruli$^{l}$, R.~Nicolaidis$^{e,f}$, F.~Nozzoli$^{e,f}$, A.~Olinto$^{u}$, D.~Orlandi$^{b}$, G.~Osteria$^{o}$, P.~A.~Palmieri$^{g,h}$, B.~Panico$^{o,p}$, G.~Panzarini$^{i,j}$, D.~Pattanaik$^{a,b}$, L.~Perrone$^{v,w}$, H.~Pessoa~Lima$^{a,b}$, R.~Pillera$^{i,j}$, R.~Rando$^{s,t}$, A.~Rivetti$^{h}$, V.~Rizi$^{k,b}$, A.~Roy$^{a,b}$, F.~Salamida$^{k,b}$, R.~Sarkar$^{a,b}$, P.~Savina$^{a,b}$, V.~Scherini$^{v,w}$, V.~Scotti$^{o,p}$, D.~Serini$^{i}$, D.~Shledewitz$^{e,f}$, I.~Siddique$^{a,b}$, L.~Silveri$^{c,d}$, A.~Smirnov$^{a,b}$, R.~A.~Torres~Saavedra$^{a,b}$, C.~Trimarelli$^{a,b}$, P.~Zuccon$^{e,f}$, S.~C.~Zugravel$^{h}$.

 \vspace{5ex}

\begin{tabular}{c}
$^{a}$ Gran Sasso Science Institute (GSSI);\\ 
$^{b}$ Istituto Nazionale di Fisica Nucleare (INFN) - Laboratori Nazionali del Gran Sasso;\\ 
$^{c}$ Center for Astrophysics and Space Science (CASS);\\ 
$^{d}$ New York University Abu Dhabi, UAE;\\ 
$^{e}$ Dipartimento di Fisica - Università di Trento;\\ 
$^{f}$ Istituto Nazionale di Fisica Nucleare (INFN) - Sezione di Trento;\\ 
$^{g}$ Dipartimento di Fisica - Università di Torino;\\ 
$^{h}$ Istituto Nazionale di Fisica Nucleare (INFN) - Sezione di Torino;\\ 
$^{i}$ Istituto Nazionale di Fisica Nucleare (INFN) - Sezione di Bari;\\ 
$^{j}$ Dipartimento di Fisica M. Merlin dell’ Università e del Politecnico di Bari;\\ 
$^{k}$ Dipartimento di Scienze Fisiche e Chimiche -Università degli Studi di L’Aquila;\\ 
$^{l}$ Départment de Physique Nuclèaire et Corpuscolaire - Université de Genève, Faculté de Science;\\ 
$^{m}$ Department of Physics and Astronomy and Astrophysics, Institute for Gravitation and the Cosmos;\\ 
$^{n}$ Department of Astronomy and Astrophysics, University of Columbia;\\ 
$^{o}$ Istituto Nazionale di Fisica Nucleare (INFN) - Sezione di Napoli;\\ 
$^{p}$ Dipartimento di Fisica E. Pancini - Università di Napoli Federico II;\\ 
$^{q}$ now at The School of Physics, The University of Sydney;\\ 
$^{r}$ CRESST/NASA Goddard Space Flight Center;\\ 
$^{s}$ Dipartimento di Fisica e Astronomia - Università di Padova;\\ 
$^{t}$ Istituto Nazionale di Fisica Nucleare (INFN) - Sezione di Padova;\\ 
$^{u}$ Columbia University, Columbia Astrophysics Laboratory;\\ 
$^{v}$ Dipartimento di Matematica e Fisica “E. De Giorgi” - Università del Salento;\\ 
$^{w}$ Istituto Nazionale di Fisica Nucleare (INFN) - Sezione di Lecce. \\
$^{z}$ On a fellowship of excellence of the Swiss Confederation.\\ 
\end{tabular}

\end{center}

\end{document}